\documentclass[twocolumn,         
               showpacs,            
               preprintnumbers,     
               aps,                 
               prd,          	    
               a4paper,             
               superscriptaddress,      
               nofootinbib,         
               tightenlines,        
               floats,floatfix,11pt
               ]{revtex4-2}              
\usepackage{graphicx}  
\usepackage[margin=0.6in]{geometry}
\usepackage{dcolumn}   
\usepackage{bm}  
\usepackage{caption}
\usepackage[normalem]{ulem}
\usepackage{subcaption}
\usepackage[utf8]{inputenc}
\usepackage{amsmath,amssymb}

\usepackage{soul}
\usepackage{float}
\usepackage{xcolor,soul}
\usepackage{makecell}
\usepackage[thinlines]{easytable}
\usepackage{array,booktabs}
\usepackage{lipsum} 
\usepackage[colorlinks=true,linkcolor=blue,citecolor=blue]{hyperref}

\usepackage{subcaption}

\begin{document}

\title{Interacting dark sector: a dynamical system perspective}
 \author{ Chonticha Kritpetch}\email{chonticha.kr@up.ac.th}
  \affiliation{High Energy Physics and Cosmology Research Group,
School of Science, University of Phayao, Phayao 56000, Thailand}
\author{Nandan Roy} \email{nandan.roy@mahidol.ac.th 
 (Corresponding Author)}
  \affiliation{NAS, Centre for Theoretical Physics \& Natural Philosophy,\\ Mahidol University,
Nakhonsawan Campus, Phayuha Khiri, Nakhonsawan 60130, Thailand}

 \author{Narayan Banerjee}\email{narayan@iiserkol.ac.in}
  \affiliation{Department of Physical Sciences, Indian Institute of Science
Education and Research Kolkata, Mohanpur 741246,WB, India}

\begin{abstract}
    We investigate the interaction between the dark sectors from the point of view of a dynamical system analysis. A general setup for interacting dark energy models that incorporates both quintessence and phantom fields through a switch
    parameter, allowing an interaction in the dark sectors, has been considered. In the first part of our analysis, we have not assumed any specific form of the interaction, and in the second part, we invoked examples in a general framework of the interaction. The potentials of the 
    scalar field are classified into two broad classes of potentials: exponential and non-exponential. We identify the potential late-time attractors of the system, which have a complete dark energy domination. From our analysis, it is evident there could be an interaction between the dark sector. The interaction, if any, weakens over time. We find for the quintessence field the transfer of energy  
    from dark matter to dark energy can flip the direction, and on the contrary, for the phantom field, it is only from dark matter to dark energy.
\end{abstract}
\maketitle

\section{Introduction}
 For the past two decades, various cosmological observations have provided substantial evidence of a universe expanding at an accelerated rate \cite{SupernovaSearchTeam:1998fmf, SupernovaCosmologyProject:1998vns, Meszaros:2002np, 
 Planck:2014loa, ahn2012ninth}; however, the explanation of this behavior still remains a challenge. Although the cosmological constant is the simplest and most successful candidate for dark energy, which drives the acceleration of the universe's expansion, it faces significant challenges. For instance, one issue is the vast difference between its required observational value and the theoretically predicted value. Another concern is the so-called coincidence problem \cite{padmanabhan2006dark}. Recent high-precision cosmological data have revealed a statistically significant discrepancy in the estimated current value of the Hubble parameter ($H_0$) when comparing early-time observations to late-time measurements. This tension presents a new challenge to the cosmological constant and signifies an open problem in cosmology. Data from the early universe measurement estimate $H_0 \sim (67.0 - 68.5)$ km/s/Mpc~\cite{Aghanim:2018eyx, Alam_2017, Joudaki:2019pmv}, while the measurement of the $H_0$ observing the local 
 universe using the distance ladder measurements 
 reported $H_0 = (74.03 \pm 1.42)$ km/s/Mpc~\cite{Riess_2022, Riess:2019cxk, Wong:2019kwg,Freedman:2019jwv}. This tension in the measurement of $H_0$ indicates the possibility of the involvement of new physics during the evolution of the universe.

Dynamical dark energy models are considered as alternatives to the cosmological constant, in order to address the challenges faced by the $\Lambda$CDM model. Various models, including quintessence, k-essence, and phantom dark energy, have been proposed \cite{amendola2010dark, Bamba:2012cp}. Generally, in these models, the accelerated expansion of the universe is driven by a scalar field rolling through a potential, creating an effective negative pressure \cite{copeland2006dynamics, Peebles2003, Armendariz2001, roy2022quintessence, Banerjee:2020xcn, Lee:2022cyh, Krishnan:2020vaf}. Although it is common to assume that dark energy evolves without any non-gravitational interactions, the possibility of interaction between dark matter and dark energy remains an open question. Interacting dark energy (IDE) models initially emerged to address the cosmic coincidence problem \cite{Cai:2004dk,mangano2003coupled,Sadjadi:2006qp,Wang:2016lxa, Jesus:2020tby}. In the IDE framework, dark matter and dark energy densities are not individually conserved; they are coupled with energy and/or momentum transferring through an interaction term. Such interaction in the dark sectors can influence the overall cosmic evolution \cite{Wang:2016lxa}. Numerous studies examine the effects of this interaction on cosmological observables \cite{amendola2000coupled, farrar2004interacting, mangano2003coupled, tamanini2015phenomenological, chimento2010linear, pan2015analytic, pettorino2005extended, pettorino2008coupled}, and these models recently showed potential to mitigate the $H_0$ and $\sigma_8$ tensions \cite{Salvatelli2014,Costa2017,DiValentino2019,Kumar2020,DiValentino:2019ffd,DiValentino:2019jae,Yang:2018euj,Wang:2018duq}. The dynamics of IDE models have been analyzed using both cosmological observations and dynamical systems analysis \cite{Salvatelli2014,Costa2017,DiValentino2019,Kumar2020,DiValentino:2019ffd,DiValentino:2019jae,Yang:2018euj,Wang:2018duq,Khyllep:2021wjd,Caldera-Cabral:2008yyo,Amendola:1999er,Boehmer:2008av,Zonunmawia:2017ofc,Hussain:2022dhp,Bahamonde:2017ize}. Dynamical systems analysis provides a qualitative framework to study non-linear systems, widely utilized to examine the stability and late-time behavior of various interacting dark energy models \cite{Khyllep:2021wjd,Caldera-Cabral:2008yyo,Amendola:1999er,Boehmer:2008av,Zonunmawia:2017ofc,Hussain:2022dhp,Bahamonde:2017ize}. The lack of consensus on interaction forms has led to the proposal of various types of interactions for studying the phenomenology and phase space behavior of IDE models. Each specific choice of interaction results in a distinct phenomenology and cosmic evolution. Various interacting quintessence or phantom dark energy models have also been studied, considering different forms of the scalar field potentials. For example, in \cite{Roy:2018eug} the interaction term has been considererd as $Q = \beta H \dot{\phi}^2$ and $Q = \frac{\beta (3H^2 - \kappa \rho_{\phi}) \dot{\phi}^2}{H}$ for a generic setup of quintessence potential. In \cite{Roy:2023uhc}, four different types of interaction have been compared against observational data, and a similar statistical preference for each of them has been indicated. The dynamics of the quintessence field using dynamical system analysis with an interaction term $Q \propto \rho_{\text{DM}} C(\phi) \dot{\phi}$ has been given in \cite{Potting:2021bje}.

Interacting phantom dark energy using dynamical system analysis has been discussed in \cite{Guo:2004xx,Paliathanasis:2024jxo}. Also, the phantom scalar field with the interaction form $Q=\Gamma \rho_m$ and $Q= \beta H \rho_m$ has been studied \cite{Halder:2024aan} for exponential and hyperbolic potentials, indicating the existance of some new type of scaling solutions.

 In this work, we aim to investigate the phase space behavior of interacting dark energy (IDE) models while keeping the choices of interaction terms and potentials as general as possible. This approach helps us understand the qualitative phase space dynamics of these models in a broader context. Our setup considers both the quintessence and phantom scalar fields within a unified framework. The potentials are classified into two broad categories: exponential forms and non-exponential forms. A detailed phase space analysis has been conducted to identify the fixed points, determine their stability properties, and identify possible late-time attractors. To test our approach, we propose a general class of interactions that can encompass a wide range of interaction forms suggested in the literature. We perform both analytical and numerical investigations to assess the stability of the system, track its evolution, and compare the model with observational data sets.
 
 The key result of this work is that the models typically converge to a final dark energy-dominated scenario. Any non-gravitational interaction, if present, diminishes over the course of evolution. Notably, in the quintessence case, the direction of energy transfer due to the interaction may reverse at some point during the evolution, whereas such a flip does not occur in the phantom models.

The paper is organized in the following way: in Sec.II, the mathematical setup of the system is discussed. Sec.III deals with the investigation of the system with exponential potential, and Sec.IV deals with the investigation of the non-exponential potential. In Sec. V, we compare our findings with previous results, and in Sec. VI, we summarize our results and findings.

\section{Mathematical Setup}

Let us consider a universe that is spatially flat and all the components of the universe follow the barotropic relation given by $p_i = w_i \rho_i$, where $p_i$ represents the pressure, $\rho_i$ represents the density, and $w_i$ represents the equation of state (EoS) of a component. 

In such a universe,  if the dark energy is considered to be a scalar field, the Einstein field equations can be written as follows:

\begin{equation} 
3 H^2=\kappa^2\left(\sum_i \rho_i+\frac{1}{2} \epsilon \dot{\phi}^2+V(\phi)\right),
\end{equation}

\begin{equation}
 \dot{H}=-\frac{\kappa^2}{2}\left(\sum_i \left(\rho_i+p_i\right)+\epsilon \dot{\phi}^2\right).
\end{equation}

We introduce the switch parameter $\epsilon$ to incorporate both quintessence and phantom field dynamics within a single framework. The equation corresponds to a quintessence field when $\epsilon = +1$ and a phantom field
when $\epsilon = -1$. In a scenario where  dark matter and dark energy interact with each other such that the density of each component is conserved together but not individually, the continuity equation for each component 
can be written as follows,

\begin{align}
& \dot{\rho}_m+3 H \rho_m=-Q,\\
& \dot{\rho}_\phi+3 H\left(\rho_\phi+p_\phi\right)=Q,
\end{align}

 where the components of dark matter and dark energy are identified by subscripts $m$ and $\phi$, respectively. The interaction term is denoted by $Q$. If $Q$ is positive, the transfer of energy happens from dark matter to dark energy and vice versa.

In this interacting dark sector scenario, the Klein-Gordon equation of the scalar field can be written as;

\begin{equation}
    \ddot{\phi}+3 H \dot{\phi}+\epsilon \frac{d V}{d \phi}= \epsilon \frac{Q}{\dot{\phi}}.
\end{equation}

To perform a dynamical system analysis, we introduce the following set of dimensionless variables,

\begin{equation}
x^2=\frac{\kappa^2 \dot{\phi}^2}{6 H^2}, y^2=\frac{\kappa^2 V(\phi)}{3 H^2}, \lambda=-\frac{1}{V(\phi)} \frac{d V(\phi)}{d \phi}.   
\end{equation}

These transformations were first introduced in \cite{Copeland:1997et} where $\lambda$ represents the steepness of the potentials.

With these, the system reduces to the following set of autonomous equations,

\begin{subequations}\label{eq:auto}
\begin{align} 
& x^{\prime}=-3 x+\sqrt{3 / 2} \epsilon \lambda y^2+\frac{3}{2} x\left(1+\epsilon x^2-y^2\right)+\epsilon  f(x, y),  \\
& y^{\prime}=-\sqrt{3 / 2} \lambda x y+\frac{3}{2} y\left(1+\epsilon x^2-y^2\right),   \\
& \lambda^{\prime}=-\sqrt{6} \lambda^2(\Gamma-1) x \label{eq:lam},
\end{align}
\end{subequations}
where $\Gamma=\frac{V(\phi) \frac{\partial^2 V(\phi)}{\partial \phi^2}}{\left(\frac{\partial V(\phi)}{\partial \phi}\right)^2}$ and a prime represent the derivative with respect to $N = \ln a$. In principle, the interaction term $Q$ can be a function of different variables 
${Q} = {Q}(\rho_m, \rho_\phi, \dot{\phi}, H, t, \ldots)$. In this work, we consider a general form of the interaction term as $Q=\sqrt{6} \dot{\phi} H^2 f(x,y)$, where $f(x,y)$ 
represents various functions involving the dynamical variables $x$ and $y$. This formulation allows for the incorporation of various types of interactions studied in cosmology. Different choices of the interaction term $Q$ result in various forms of the function $f(x,y)$. For a comprehensive list of different choices of $Q$ with corresponding $f(x,y)$, refer to Table 16 in \cite{Bahamonde:2017ize} and the references therein. From the physical point of view, in a purely dark matter or dark energy-dominated state, the interaction and its derivatives should vanish; that is  
$f(x,y)= 0 = \frac{\partial f(x,y)}{\partial x} = \frac{\partial f(x,y)}{\partial y}$. The dynamical system variables can be used to express various cosmological parameters concisely as,
\begin{align}
    &\Omega_\phi = \epsilon x^2 + y^2, \\
    &w_\phi = \frac{\epsilon x^2 - y^2}{\epsilon x^2 + y^2},\\
    &q= -1 + \frac{3}{2}(1+\epsilon x^2 - y^2).
\end{align}

Here $\Omega_\phi = \frac{1}{2} \epsilon \dot{\phi}^2 + V(\phi)$ represents the density parameter of the scalar field, $w_\phi$ represents the equation of state parameter of the scalar field, and $q$ is the deceleration parameter of the universe.

To close the autonomous system given in equation (\ref{eq:auto}), it is necessary to specify a particular form of the $\Gamma$ function, which can be essentially related to choosing a specific form of the potential. In general, $\Gamma$ can be evolving. It can be noticed from Eq.(\ref{eq:lam}) that depending on the choice of the $\Gamma$, we can classify the system into two classes. The first class arises when $\Gamma = 1$, resulting in an exponential potential that effectively reduces the system to a 2-dimensional form. On the other hand, the second class corresponds to $\Gamma \neq 1$, which corresponds to all potentials except the exponential potential. This classification has already been used in \cite{Roy:2017mnz, Roy:2014hsa}.

In the next section, we discuss the fixed points of the system and the corresponding stability of those fixed points. Fixed points are obtained by simultaneously solving the autonomous equations given in Eq.(\ref{eq:auto})  
with $x^\prime = y^\prime = \lambda^\prime = 0$. A fixed point is said to be stable if all the eigenvalues of the Jacobian matrix at that fixed point have negative real parts and unstable if all of them have a positive real part. The point is a saddle if at least one eigenvalue has a positive real part and one has a negative real part.

\section{Exponential Potential}\label{sec:expo}

In this case, we assume that the potential has an exponential form and set $\Gamma = 1$. This reduces the dimension of the system from 3D to 2D since $\lambda$ is constant.

\subsection{Fixed point and stability}
The fixed points of this system are given in Table \ref{tab:fixedpoints_exp}, and in the last column, we tabulated the form of the interaction $f(x,y)$ at the fixed points. There are only three classes of fixed points.

\begin{table*}[]
\centering
\resizebox{\textwidth}{!}{%
\begin{tabular}{|c|c|c|c|}
\hline
Fixed Points & $x$                                                                                                                    & $y$                       & $f(x,y)$                                                                                                                                \\ \hline
$Q_1$        & 0                                                                                                                      & 0                        & 0                                                                                                                                       \\ \hline
$Q_{2i}$        & \begin{tabular}[c]{@{}c@{}}$\frac{3}{2} x (1+\epsilon x^2)- 3 x  + \epsilon  f(x,0)=0$\end{tabular}                                                   & 0          & \begin{tabular}[c]{@{}c@{}}$f(x,y) = 0 $ for $x=0,x\pm1$ \\ $f(x,y)\neq 0 $ for $0<x<1$ \end{tabular} \\ \hline
$Q_{3 i}$        & $\frac{1}{2} \left(\epsilon  \left(2 f(x,y)+\sqrt{6} \lambda \right)+2 \sqrt{6} \lambda  x^2 -2 x \left(\lambda ^2 \epsilon +3\right)\right) =0$                                                                                                                      & $\pm \frac{\sqrt{3 x^2 \epsilon -\sqrt{6} \lambda  x+3}}{\sqrt{3}}$                        & $f(x,y)$                                                                                                                     \\ \hline

\end{tabular}}
\caption{List of the fixed points and the corresponding $f(x,y)$ for the exponential potential with $\Gamma =1$.}
\label{tab:fixedpoints_exp}
\end{table*}

\subsubsection{Fixed Point $Q_1$}
The fixed point $Q_1$ corresponds to a completely dark matter-dominated ($\Omega_\phi =0$) regime, and the interaction term $f(x,y)$ vanishes at this fixed point. The eigenvalues associated with $Q_1$ are given by ($\frac{3}{2}, -\frac{3}{2}$). This fixed point is a saddle in nature. Since the deceleration parameter $q = \frac{1}{2}$ at this fixed point, it corresponds to a decelerated expansion of the universe.

\subsubsection{Fixed Point $Q_2$}
Depending on the choice of the function $f(x,y)$, the class of the fixed points $Q_2$ can be a single or multiple fixed points upon solving the equation given in Table \ref{tab:fixedpoints_exp},

\begin{equation} \label{eq:Q2}
\frac{3}{2} x (1+\epsilon x^2) - 3 x + \epsilon f(x,0)=0
\end{equation}

The eigenvalues associated with this fixed point are given by  $\frac{1}{2} \left(3 x^2 \epsilon -\sqrt{6} \lambda  x+3\right),\frac{1}{2} \left(2 \epsilon  \partial_x f+9 x^2 \epsilon -3\right)$. For a completely matter-dominated 
fixed point ($\Omega_m =1$)  for which $x=0$, one gets back the $Q_1$ as the fixed point. For the complete kinetic energy domination of the quintessence field, $(x=\pm 1)$ these fixed points are unstable fixed points since the eigenvalues reduce 
to $[\frac{1}{2}(6 \pm \sqrt{6} \lambda), 3]$. For the phantom field, the complete kinetic energy domination fixed point does not exist since $x^2 = 1$ does not satisfy the Eq.(\ref{eq:Q2}).

This fixed point could also represent the scenarios where both dark matter and dark energy contribute; the stability conditions are as follows. For the quintessence field ($\epsilon = +1$),

\begin{subequations} \label{eq:expq2qf}
\begin{align}
\left(-1\leq x<0 ; \lambda <\frac{3 x^2+3}{\sqrt{6} x} ; \partial_x f<\frac{1}{2} \left(3-9 x^2\right)\right) ,  \\
\left(0<x\leq 1; \lambda >\frac{3 x^2+3}{\sqrt{6} x}; \partial_x f<\frac{1}{2} \left(3-9 x^2\right)\right),
\end{align}
\end{subequations}

and for the phantom field ($\epsilon = -1$),
\begin{subequations}\label{eq:expq2ff}    
\begin{align}
 \left(x<0; \lambda <\frac{3-3 x^2}{\sqrt{6} x}; \partial_x f>\frac{1}{2} \left(-9 x^2-3\right)\right) ,\\
 \left(x>0; \lambda >\frac{3-3 x^2}{\sqrt{6} x}; \partial_x f>\frac{1}{2} \left(-9 x^2-3\right)\right) .  
\end{align}
\end{subequations}

Contrary to the canonical field ($0 \leq \Omega_{\phi} \leq 1$), there is no strict positivity condition on the energy density of the phantom field. It can be negative ($\Omega_{\phi} < 0$) too. These fixed points have previously been reported in \cite{Urena-Lopez:2005pzi,Gumjudpai:2005ry}. Since $y=0$ at this fixed point, the equation of state of both the scalar field is $w_\phi = 1$. Hence, the scalar fields behave as a stiff fluid. This point corresponds to a decelerated expansion for both fields since $q = \frac{1}{2} + \frac{3}{2} w_\phi \Omega_\phi > 0$ for $0 \leq \Omega_\phi \leq 1$.

\subsubsection{Fixed Point $Q_3$}
The general form of the class of fixed points $Q_3$ is given in the form of an equation. 
\begin{equation}
    \frac{1}{2} \left(\epsilon  \left(2 f(x,y)+\sqrt{6} \lambda \right)+2 \sqrt{6} \lambda  x^2 -2 x \left(\lambda ^2 \epsilon +3\right)\right) = 0.
\end{equation}

One can solve the above algebraic equation for a given form of the interaction $f(x,y)$ to find all associated fixed points. Although the eigenvalues corresponding to this point can be quite complicated in form, for a 2D system, one can 
use the trace and determinant of the Jacobian matrix to investigate the stability. The trace $T_{Q3}$ and determinant $D_{Q3}$ of the Jacobian matrix at these fixed points are given in Appendix \ref{appen:eigenQ3}. The condition for the stability of this fixed point is $T_{Q3} < 0$ and $D_{Q3} > 0$.

As an example, let us consider a special case where $x=0$ and $y=\pm1$, depicting a completely dark energy-dominated universe and hence $f(x,y)=0$ and $\partial_x f=\partial_y f=0$, the trace and determinant of the fixed point reduce to the following simple form:

\begin{align}
T_{Q3}&= -6,  \\
D_{Q3}&=  3  \epsilon \lambda ^2  +9.
\end{align}

For the quintessence field($\epsilon = +1$), the fixed point in this example is stable. On the contrary for the phantom field ($\epsilon = -1$), it depends on the choice of $\lambda$. Since we have chosen $x=0$ and $y=\pm1$, the field has a slow roll, and $\lambda<<1$. Therefore, even for the phantom field, this fixed point is an attractor. At this fixed point, the EoS of the scalar field and the deceleration parameter of the universe depend on the choice of the particular form of the interaction. However, for the choice of $x=0, y=\pm1$ the deceleration parameter and EoS are respectively $q=-1, w_\phi = -1$ indicating an accelerating universe. 

These fixed points can represent the scaling solutions or the complete dark energy domination, which was already reported in the literature \cite{Gumjudpai:2005ry,Copeland:1997et,Amendola:2020ldb}.

\subsection{Example}

In order to investigate the general setup further, here, we propose a general form of interaction as an example to test our approach:

\begin{equation} \label{eq:int_example}
  f(x,y) = \alpha  (1 - \epsilon x^2 -y^2)^m x^\gamma   
 \end{equation}

This particular form of interaction allows us to examine a wide range of interactions \cite{Bahamonde:2017ize,Gumjudpai:2005ry,Boehmer:2008av,Chen:2008pz,Mimoso:2005bv}. In Table \ref{tab:interactions} of Appendix \ref{appen:interactions}, we have given a list of interactions that are used in literature and can be incorporated into the above general form. It is important to note that this form is not limited to only those specific interactions mentioned in Table \ref{tab:interactions}.

For this choice of the $f(x,y)$, the fixed point $Q_1$ exists only when $\gamma > 0$, as the condition $f(x,y) = 0$ is necessary for the existence of this fixed point. Any choice of interaction that violates this criterion would miss the fixed point representing a pure matter-dominated universe that does not exist. Furthermore, this fixed point is inherently unstable for both the quintessence and phantom fields, regardless of the specific form of the interaction chosen.

The particular fixed points included in $Q_2$ class can be found by solving the quadratic equation;

\begin{equation}\label{eq:Q22}
    \frac{3}{2} x (1+\epsilon x^2) + \epsilon \alpha (1-\epsilon x^2)^m x^\gamma - 3x =0. 
\end{equation}

For both the quintessence and phantom case, there are multiple solutions to the above equations leading to multiple fixed points, with $x=0$ being the trivial solution. The fixed point $x=0$ is indistinguishable from the 
matter-dominated case $Q_1$.   It has already been shown in the general analysis in the previous subsection that for the complete quintessence field kinetic domination ($x^2 = 1$), these fixed points are stable and for the phantom field kinetic domination, they do not exist. 

 
 Depending on the choice of $\gamma, m$, there could be fixed points that can represent a state of the universe where there are both dark matter and dark energy contributions. From our general analysis in the previous 
 subsection, these points could be unstable or saddle in nature. For the quintessence field, this particular interaction renders the stability conditions in the expressions (\ref{eq:expq2qf}) to the following,

 \begin{align}
\left(-1\leq x<0 ; \lambda <\frac{3 x^2+3}{\sqrt{6} x} ; f<\frac{3x (1-x^2)(1-3x^2)}{\gamma (1-x^2) - 2 x^2} \right)  \\
\left(0<x\leq 1; \lambda >\frac{3 x^2+3}{\sqrt{6} x}; f<\frac{3x (1-x^2)(1-3x^2)}{\gamma (1-x^2) - 2 x^2} \right).
\end{align}

Similarly, for the phantom field, conditions in the expressions (\ref{eq:expq2ff}) reduce to;

\begin{align}
 \left(x<0; \lambda <\frac{3-3 x^2}{\sqrt{6} x}; f>-\frac{3x (1+x^2)(1+3x^2)}{\gamma (1+x^2) +2 x^2}\right) \\
 \left(x>0; \lambda >\frac{3-3 x^2}{\sqrt{6} x}; f>-\frac{3x (1+x^2)(1+3x^2)}{\gamma (1+x^2) +2 x^2}\right)   
\end{align}.

Here we have used $\partial_x f(x,y)= f(x,y) (\frac{\gamma}{x} - \frac{2 \epsilon x}{1 - \epsilon x^2 -y^2})$ and $y=0$ at this fixed point.

To obtain all the fixed points belonging to this $Q_3$ class, one needs to solve the equation,

\begin{equation}
\begin{split}
    \frac{1}{2} \Biggl(\epsilon  \left(\sqrt{6} \lambda +2 \alpha  x^{\gamma } \left(1-x^2 \epsilon \right)^m\right)  \\ 
    +2 \sqrt{6} \lambda  x^2-2 x \left(\lambda ^2 \epsilon +3\right)\Biggr)=0
\end{split}
\end{equation}

There could be multiple fixed points depending on the choice of $\gamma, m$. One can easily compute the trace and determinant given in Appendix \ref{appen:eigenQ3} and find the stability of these fixed points. We then use numerical techniques to find the phase space behaviour and evolution of the system for different choices of the model parameters.

\begin{figure*}[!h]
    \centering    \includegraphics[width=\textwidth]{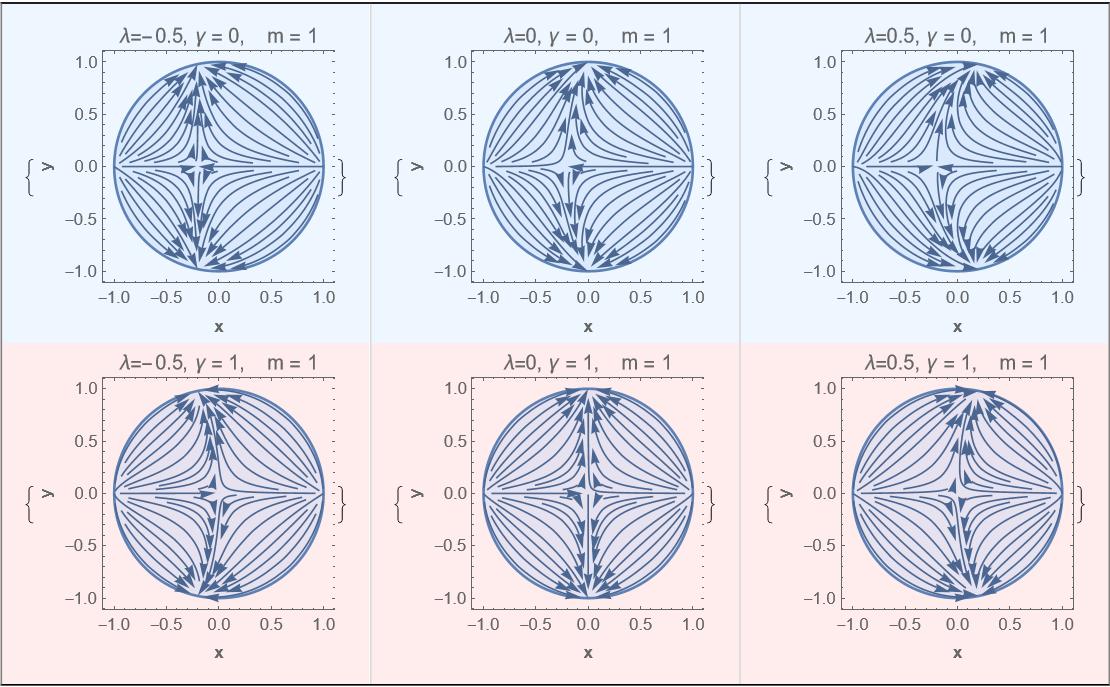}
    \caption{Phase plot on the $x$ vs $y$ plane for the exponential potential ($\Gamma = 1$) with $\alpha = -0.2$ for different choices of $\lambda$, $\gamma$, and $m$ for the quintessence field. The shaded circle represents the Friedmann constraint with $0 \leq x^2 + y^2 \leq 1$.}
    \label{fig:phase-exp}
    
\end{figure*}

In Fig.~\ref{fig:phase-exp}, we have shown the phase plot of the system with the exponential potential for different choices of the $\lambda, \gamma$ and $m$ parameters for the quintessence field. Here, we have considered $\alpha=-0.2$. 
Our choices of the $\alpha$ parameter are motivated by the posterior obtained for the interaction parameter $\alpha$ using a similar mathematical setup in \cite{Roy:2023uhc}. The first row (blue background) and the second row (orange background) represent $\gamma=0$ and $\gamma = 1$, respectively. For these plots, we have considered $m=1$ and $\lambda = [-0.5,0,+0.5]$. The circle in the plots represents the Friedman constraint for the quintessence field, defined by $0 \leq \Omega_\phi \leq 1$. The boundary of this circle signifies a completely dark energy-dominated state of the universe. In all these plots, the late-time attractors are located at the boundary of the circle, indicating that a completely dark energy-dominated universe is a late-time attractor. Notice that with the change in the sign of $\lambda$ the sign of the value of the $x$ at the late time attractor also changes. Positive $\lambda$ corresponds to a positive value of $x$ and vice versa. 

Also, these plots agree with the analytical finding that for $\gamma =0$ case, there would not be a fixed point that is completely dominated by the dark matter component. The first row corresponds to the $\gamma = 0$ case with no fixed points at $x=0, y=0$. On the other hand, the second row where $\gamma =1$, $x=0, y=0$ clearly seem to have a saddle fixed point. 

\begin{figure*}[h!]
    \centering   \includegraphics[width=\linewidth]{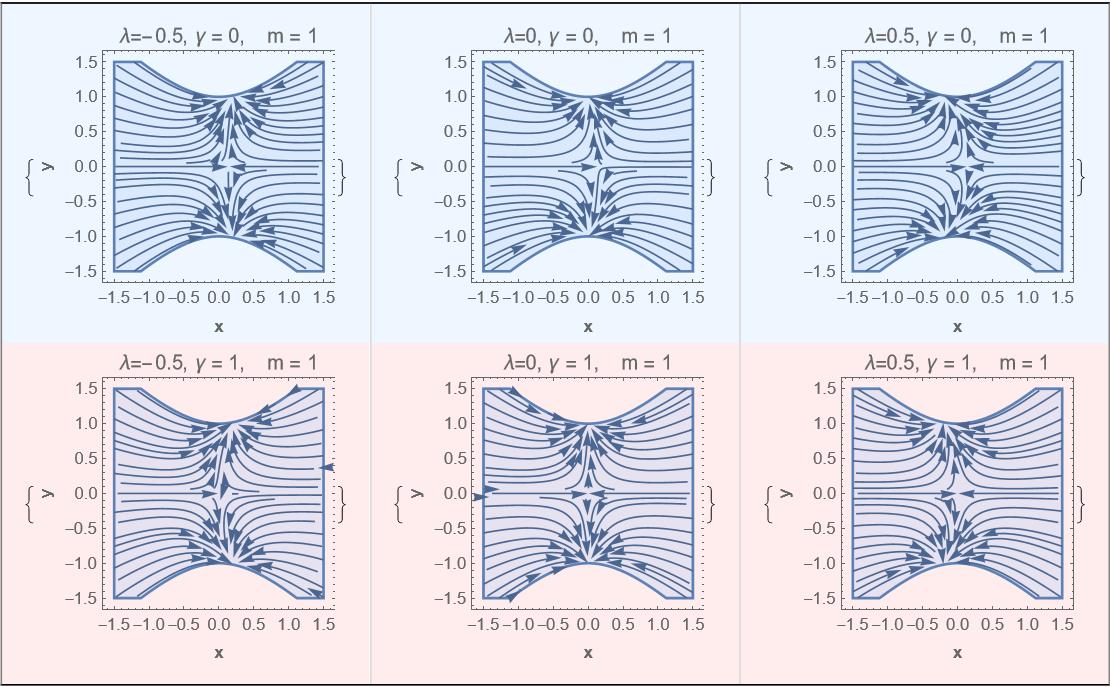}
    \caption{Phase plot on the $x$ vs $y$ plane for the exponential potential ($\Gamma = 1$) with $\alpha = -0.2$ for different choices of $\lambda$, $\gamma$, and $m$ for the phantom field. The shaded hyperbolic region represents the Friedmann constraint with $0 \leq  -x^2 + y^2 \leq 1$.}
    \label{fig:phase-exp-phant}
\end{figure*}

For the phantom field in Fig.~\ref{fig:phase-exp-phant} we have shown the phase plot of the system for the same choice of the parameters as in the quintessence case. The region 
 $ 0 \leq \Omega_\phi \leq 1$ which is represented by the shaded hyperbola. In this case, the phantom cases have only two classes of fixed points as those belonging to the $Q_{2i}$ coincide with the $Q_1$.  Similar to the quintessence case, the phantom field also shows complete dark energy domination as the late-time attractor. This is because the stable fixed points are located on the boundary of the hyperbola. For the phantom field, it is possible for trajectories that originate outside the region $0 \leq \Omega_\phi \leq 1$ to eventually enter this region, as there are no attractors outside it. This similar behavior of the phantom field has been reported in \cite{Urena-Lopez:2005pzi}.

\begin{figure*}[h!]
     \centering
     \begin{subfigure}[b]{0.49\textwidth}
         \centering
         \includegraphics[width=\textwidth]{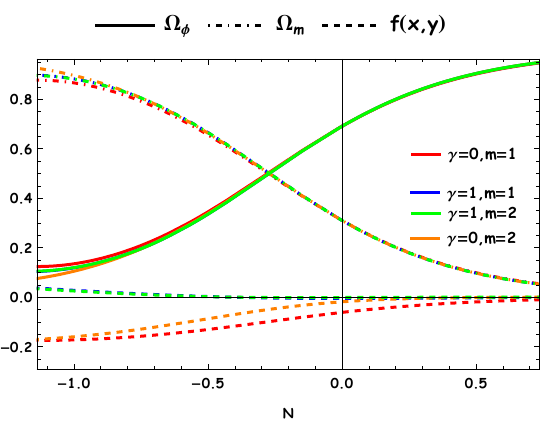}
         \caption{The plot of the evolution of the density parameter $\Omega_\phi$ (in solid) and $\Omega_m$ (in dot-dashed) and the interaction $f(x,y)$ (in dashed) for the quintessence field with exponential potential.}
         \label{fig:exp_om_q}
     \end{subfigure}
     \hfill
     \begin{subfigure}[b]{0.49\textwidth}
         \centering
         \includegraphics[width=\textwidth]{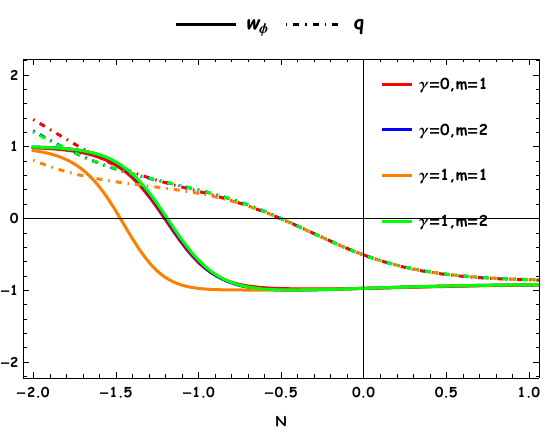}
         \caption{The plot of the evolution of the equation of state of the scalar field $w_\phi$ (in solid) and the deceleration parameter $q$ (in dashed) for the quintessence field with exponential potential. }
         \label{fig:exp_w_q}
     \end{subfigure}
     \hfill
      \begin{subfigure}[b]{0.49\textwidth}
         \centering
         \includegraphics[height=0.8\textwidth,width=1.\textwidth]{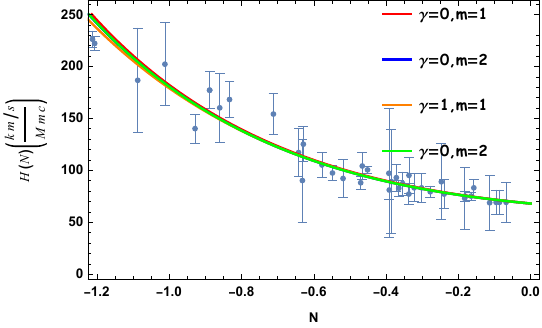}
         \caption{Plot of the $H(N)$ vs $N$ together with the observational data for the comparison of the quintessence field with exponential potential.}
         \label{fig:exp_H_q}
     \end{subfigure}
     \hfill
     \begin{subfigure}[b]{0.49\textwidth}
         \centering
         \includegraphics[width=\textwidth]{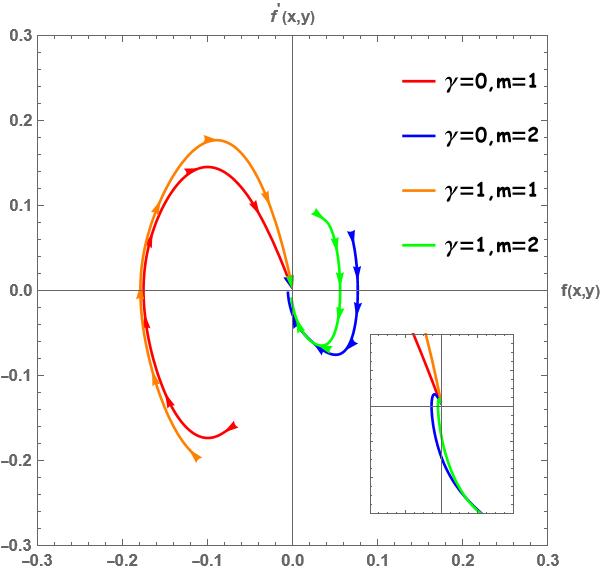}
         \caption{Phase plot displaying $f(x,y)$ against its derivative $f'(x,y)$ of the quintessence field with exponential potential. The inset showcases a zoomed-in view of the plot's central region for clarity.}
         \label{fig:exp_f_q}
     \end{subfigure}
        \caption{Plot of the different cosmological parameters and the phase space of the interaction $f(x,y)$ of the quintessence field for the exponential potential ($\Gamma =1$) with $\alpha = -0.2$.}
        \label{fig:exp_quint}
\end{figure*}

For further investigation of the evolution of the system, we have numerically evolved the autonomous system given in Eq.(\ref{eq:auto}). For the numerical solution of the system, either one needs to supply the initial condition 
or the current condition of the $x,y$. Here, we consider the second approach, where we have estimated the current values of the $x,y$ from the observation by solving, 
$\Omega_{\phi 0} = \epsilon x_0 ^2 + y_0 ^2$ and $q_0 = -1 + \frac{3}{2}(1+\epsilon x_0 ^2 - y_0 ^2 )$ where $\Omega_{\phi 0}$ and $q_0$ are the current density parameter of the scalar field and the current value of the deceleration parameter 
respectively. Here we consider $\Omega_{\phi 0} = 0.68$\cite{Aghanim:2018eyx} and $q_0 = -0.51 $\cite{Riess:2021jrx}. For quintessence field ($\epsilon=+1$) we estimate $x_0 = 0.09, y_0=0.825$ and for the phantom field 
$x_0 = 0.01, y=0.824$. In Fig.~\ref{fig:exp_quint}, we have shown the evolution of different cosmological variables. We have chosen $\lambda = 0.5,\alpha=-0.2$ for the quintessence field and $\lambda = 0.2,\alpha=-0.2$ for the phantom field. 
The choice of $\lambda$ is arbitrary, but the particular choices we have here are to avoid difficulties in the numerical integration and also to fit data. For the choice of $\alpha$ parameter, we have used the constraints obtained from recent 
cosmological observation using a similar mathematical setup\cite{Roy:2023uhc}. In Fig.~\ref{fig:exp_om_q} we have shown the evolution of the density parameters $\Omega_\phi$,$\Omega_m$ and $f(x,y)$. Evolution of the density parameters 
$\Omega_\phi$ and $\Omega_m$ have the expected behavior whereas $f(x, y)$ (in dashed line) shows some intriguing nature. In the distant past, the magnitude of the interaction $f(x, y)$ was much larger when compared to the present epoch 
and approaches zero asymptotically for the future. For all the cases, the maximum value of the interaction is during the matter domination and starts to decrease as the dark energy gradually dominates. 
To understand the evolution of the interaction better in Fig.~\ref{fig:exp_f_q}, we have presented the phase plot of the interaction term $f(x,y)$ vs $f^\prime(x,y)$ using the Eq.(\ref{eq:int_example}) and (\ref{eq:auto}). It is 
interesting to note that the evolution of the $f(x,y)$ started from a non-zero value and evolved to zero as the universe \textbf{become} completely dark energy-dominated. Another interesting fact to notice here is that, for some cases, the interaction 
term $f(x,y)$ has a flip in signature.

In  Fig.~\ref{fig:exp_w_q}, we have plotted the evolution of the EoS of the scalar field together with the deceleration parameter $q$ and in Fig.~\ref{fig:exp_H_q}, we have plotted the evolution of the Hubble parameter $H(N)$ with 
respect to $N$. For comparison with the observational data, we have also plotted cosmic chronometers data (please see Table 1 of Ref.\cite{Favale:2023lnp} ). As it is evident, these models of interaction can fit the data quite well, particularly at the late time. Recently, the full (non-diagonal) covariance matrix of the data points is computed in \cite{Moresco:2018xdr}. Since our analysis does not involve any statistical analysis, we will not be using the covariance matrix here.

\begin{figure*}[!h]
     \centering
     \begin{subfigure}[b]{0.49\textwidth}
         \centering
         \includegraphics[width=\textwidth]{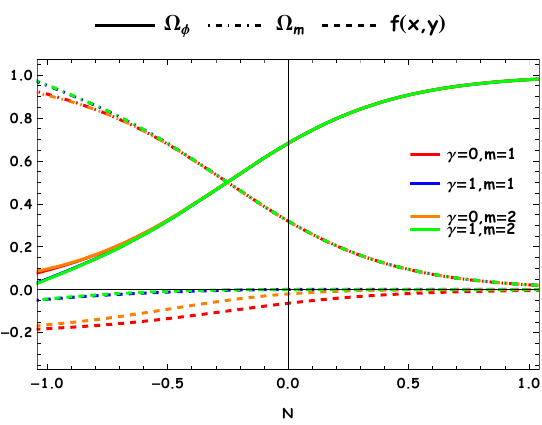}
         \caption{The plot of the evolution of the density parameter $\Omega_\phi$ (in solid) and $\Omega_m$ (in dot-dashed) and the interaction $f(x,y)$ (in dashed) of the phantom field with exponential potential.}
         \label{fig:exp_om_ph}
     \end{subfigure}
     \hfill
     \begin{subfigure}[b]{0.49\textwidth}
         \centering
         \includegraphics[width=\textwidth]{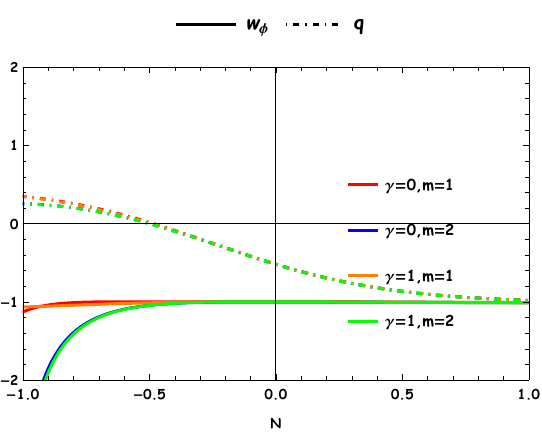}
         \caption{The plot of the evolution of the equation of state of the scalar field $w_\phi$ (in solid) and the deceleration parameter $q$ (in dashed) of the phantom field with exponential potential.}
         \label{fig:exp_w_ph}
     \end{subfigure}
     \hfill
      \begin{subfigure}[b]{0.49\textwidth}
         \centering
         \includegraphics[height=0.9\textwidth,width=1.\textwidth]{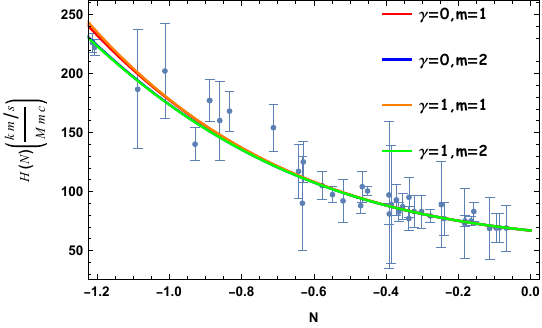}
         \caption{Plot of the $H(N)$ vs $N$ together with the observational data for the comparison of the phantom field with exponential potential.}
         \label{fig:exp_H_ph}
     \end{subfigure}
     \hfill
     \begin{subfigure}[b]{0.49\textwidth}
         \centering
         \includegraphics[width=\textwidth]{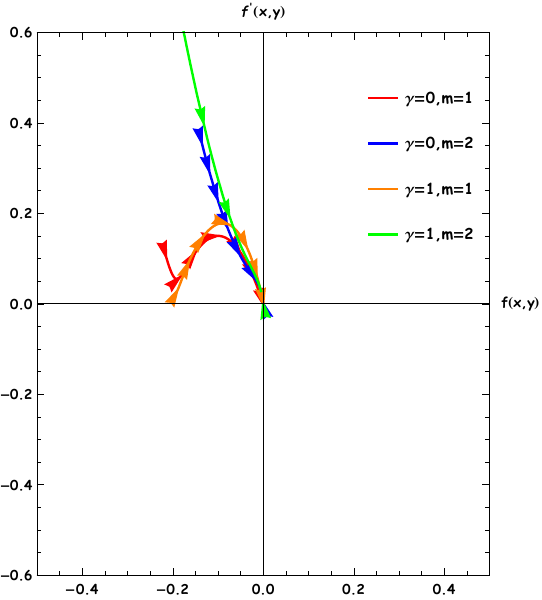}
         \caption{Phase plot displaying $f(x,y)$ against its derivative $f'(x,y)$ of the phantom field with exponential potential.}
         \label{fig:exp_f_ph}
     \end{subfigure}
        \caption{Plot of the different cosmological parameters and the phase space of the interaction $f(x,y)$ of the phantom field for exponential potential $(\Gamma =1)$ with $\alpha = -0.2$.}
        \label{fig:exp_phant}
\end{figure*}

In Fig.~\ref{fig:exp_phant}, we present the phantom case for the same choice of $\lambda$ and $\alpha$ parameters. $\Omega_m$ and $\Omega_\phi$ show the expected behavior.  The evolution of $f(x,y)$ has a behavior similar to the 
quintessence case. The interaction is found to have the maximum in magnitude during the matter domination and decrease gradually to zero in the future, which is completely dominated by the phantom field. But if we notice the evolution 
of $f(x,y)$ from 
the phase plot given in Fig.~\ref{fig:exp_f_ph}, there is a qualitative difference with the quintessence case. For all the cases we have considered the evolution of the interaction is unidirectional for the phantom case since there is no flip 
in the signature of the $f(x,y)$, and the interaction vanishes faster than the quintessence case.

The evolution of the $w_\phi$ and the deceleration parameter $q$ for the phantom field is shown in Fig.~\ref{fig:exp_f_ph}. As it is expected $w_\phi$ evolves from $w_\phi<-1$ to $w_\phi=-1$ at present. The comparison with the 
cosmic chronometers data (please see Table 1 of Ref.\cite{Favale:2023lnp} ) is shown in Fig.~\ref{fig:exp_H_ph} by plotting $H(N)$ vs $N$, and it can be seen that the data can be fitted very well even with the phantom field as dark energy.

\section{Non-Exponential Potential}\label{sec:non-expo}
Here, we present the phase space behavior of the class of potentials that are non-exponential, characterized by the condition $\Gamma \neq 1$. The fixed points associated with this class of potentials are listed in Table \ref{tab:fixedpoints}. In total, there are four classes of fixed points.

\begin{table*}[]
\centering
\resizebox{\textwidth}{!}{%
\begin{tabular}{|c|c|c|c|c|}
\hline
Fixed Points & $x$                                                                                                                    & $y$                      & $\lambda$  & $f(x,y)$                                                                                                                                \\ \hline
$P_1$        & 0                                                                                                                      & 0                        & $\lambda$  & 0                                                                                                                                       \\ \hline
$P_{2i}$        & $\frac{3}{2} x (1-\epsilon x^2)-\epsilon  f(x,0)=0$                         & 0                        & 0          & \begin{tabular}[c]{@{}c@{}}$f(x,y=0$ \\ for $x=0,\pm1$\end{tabular} \\ \hline
$P_{3\pm}$        & 0                                                                                                                      & $\pm1$                        & $\lambda=-\sqrt{\frac{2}{3}}  f(x,y)$ & $f(x,y)$                                                                                                                     \\ \hline
$P_{4\pm}$        & \begin{tabular}[c]{@{}c@{}}$3x- \epsilon  f(x,y)=0$
\end{tabular} & $y^2 = (1+\epsilon x^2)$ & 0  & $f(x,y)$ \\ \hline

\end{tabular}}
\caption{List of the fixed points for the non-exponential potentials ($\Gamma \neq 1$) with corresponding $f(x,y)$.}
\label{tab:fixedpoints}
\end{table*}

\subsubsection{Fixed Point $P_1$}
Fixed Point $P_1$ represents a completely matter-dominated situation for both the quintessence and the phantom fields ($\epsilon = \pm 1$). The corresponding eigenvalues of this fixed point are ($\frac{3}{2},0,-\frac{3}{2}$). 
Regardless of the specific form of the interaction function $f(x,y)$, this fixed point is inherently saddle in nature. Since the value of the deceleration parameter at this fixed point is $q=1/2$, it indicates a decelerated expansion of the universe.

\subsubsection{Fixed Point $P_2$}
The cosmological behaviour of the fixed point $P_2$ is similar to the fixed point $Q_2$ in the exponential case, except that $\lambda =0$ for $P_2$. The particular fixed points corresponding to this class can be obtained from Eq.(\ref{eq:Q2}). 
The eigenvalues corresponding to these fixed points are ($0,\frac{3}{2} \left( x^2 \epsilon +1\right),\frac{1}{2} \left(2   \epsilon \partial_x f+9 x^2 \epsilon -3\right)$). These fixed points are nonhyperbolic fixed points, but 
it can be easily checked that for the quintessence field ($\epsilon = +1$), the second eigenvalue cannot be negative. Hence, for the quintessence field, these fixed points are unstable.

For the phantom field, the second eigenvalue is given by $\frac{3}{2}(1 - x^2)$. This expression can also be written as $\frac{3}{2}(1 + \Omega_{\phi_{P_2}})$ by noting that $\Omega_{\phi_{P_2}} = -x^2$ at this fixed point. This fixed point is considered unstable for values of $\Omega_\phi$ in the range $-1 < \Omega_\phi \leq 1$. 

Similar to the $Q_2$ fixed point for the exponential potential case, at this fixed point, the equation of state of the scalar field is $w_\phi = 1$, hence, it behaves as a stiff fluid. It also corresponds to a decelerated expansion for both fields since $q = \frac{1}{2} + \frac{3}{2} w_\phi \Omega_\phi > 0$ for $0 \leq \Omega_\phi \leq 1$.

\subsubsection{Fixed Point $P_3$}

The fixed point $P_3$  represents a completely dark energy-dominated situation where the value of the $\lambda$ depends on the choice of the form of the interaction. Since this is a completely dark energy-dominated fixed point, from a physical 
point of view there could not be any interaction between the dark energy and the dark matter because of the absence of the latter. Hence we consider the interaction term $f(x,y) = \partial_xf(x,y) =0$, therefore $\lambda = 0$ at this 
fixed point. 

The eigenvalues corresponding to this fixed point are $(-3,0,-3)$. This is a nonhyperbolic fixed point, so one cannot use the linear stability analysis. A more complex analytical tool like the central manifold theorem or the numerical tools have to be used to analyse the stability of this fixed point. At this fixed point, the EoS of the scalar field is $w_\phi = -1$, and this fixed point corresponds to an accelerated expansion of the universe since $q = -1$.

\subsubsection{Fixed Point $P_4$}

The fixed point $P_{4\pm}$ is an intermediate fixed point for which there can be contributions from both dark energy and dark matter. The corresponding eigenvalues at this fixed point are the following:
 
$\begin{aligned} & \left\{0,\frac{1}{6} \left(-A +3 \epsilon \partial_xf-18\right),\frac{1}{6} \left(A +3 \epsilon \partial_xf-18\right)\right\},\end{aligned}$ where

\begin{equation}
\begin{split}
    A=\sqrt{3} \sqrt{4f^2(\partial_x f - 3 \epsilon) + 4 \epsilon f \sqrt{9 + \epsilon f^2} \partial_y f  + 3 (\partial_x f)^2}
\end{split}
\end{equation}

This fixed point is also nonhyperbolic, and more sophisticated numerical or mathematical methods should be used to study its stability with a particular choice of interaction. Although the value of the EoS of the scalar field depends on the choice of the interaction, it represents an accelerated expansion of the universe since $q = -1$ at this fixed point.

\subsection{Example}

For further investigation of the non-exponential potential case, here we consider the same form of the interaction given in Eq.(\ref{eq:int_example}).

 Like the exponential case, the fixed point $P_1$ only exists for $\gamma > 0$ models, and it is unstable, independent of the form of the interaction.

All the fixed points belonging to $P_2$ class can be found by solving the quadratic equation given in Eq.(\ref{eq:Q22}). Irrespective of any particular form of the interaction, this fixed point is unstable for both the quintessence and the
phantom field.

The fixed point $P_3$ exists for any choice of $\gamma$ and $m$, and it is a completely dark energy-dominated state. The eigenvalues at this fixed point are ($-3,0,3$) as $\partial_x f =0$ at this fixed point. This fixed point is 
non-hyperbolic, so its stability cannot be understood using linear stability analysis. One can investigate it numerically for some specific choices of the model parameters $\gamma, m$, and $\Gamma$.

To get the particular fixed point belonging to the $P_4$ class, one needs to solve the following equation,

\begin{equation}
    3x -  \epsilon \alpha  x^{\gamma } \left(1 -\epsilon x^2 -y^2\right)^m =0.
\end{equation}
One can notice from another equation corresponding to this fixed point $y^2 = (1 + \epsilon x^2)$, for the quintessence for any $x\neq 0, y^2 >1$. Hence, the only physical solution to this equation is $x=0$, which makes this fixed point 
indistinguishable from  $P_3$. For the phantom field, the behaviour is richer as this fixed point can represent complete matter domination to dark energy domination, and a combination of both depending on the choice of $\gamma, m$. This 
fixed point is also nonhyperbolic. In the next, we shall numerically investigate the phase space behaviour and the evolution of this system for this particular choice of interaction.

\begin{figure*}[h!]
    \centering   \includegraphics[width=\linewidth]{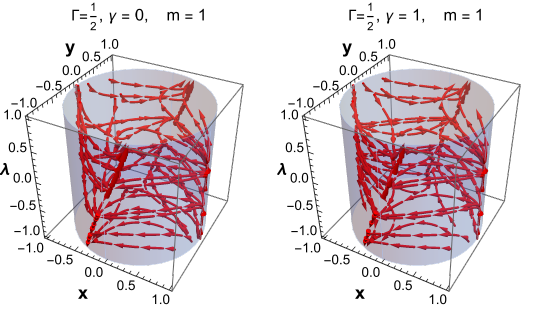}
    \caption{3D phase plot for the quintessence field for $x, y, \lambda$ for the non-exponential potential ($\Gamma = \frac{1}{2}$) with $\alpha = -0.4$ for different choices of $\gamma, m$. The Friedman constraint here is represented by the surface of the shaded cylindrical region. The surface of the cylinder corresponds to complete dark energy domination and the center represents complete matter domination.} 
    \label{fig:phase-q1}
\end{figure*}

\begin{figure*}[h!]
    \centering   \includegraphics[width=\linewidth]{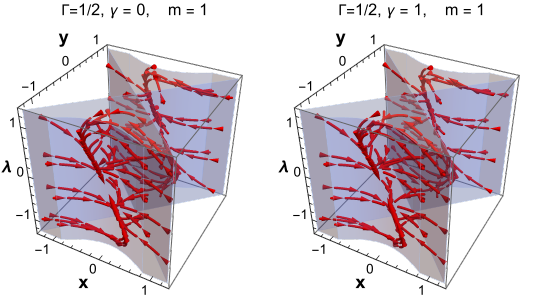}
    \caption{3D phase plot for the phantom field for $x, y, \lambda$ for the non-exponential potential ($\Gamma = \frac{1}{2}$) with $\alpha = -0.4$ for different choices of $\gamma, m$. The Friedmann constraint here is represented by the surface of the hyperboloid shown by the shaded region. The surface of the hyperboloid corresponds to a complete dark energy domination, and the center represents complete matter domination.}
    \label{fig:phase-ph1}
\end{figure*}

For the numerical investigation, we chose $\Gamma = 1/2$ for which the potential becomes $V(\phi) = (A + B \phi)^2$ and $\alpha = -0.4$.  In Fig.~\ref{fig:phase-q1} and Fig.~\ref{fig:phase-ph1}, we have plotted the phase diagram for $\Gamma = 1/2$, considering the same combination of the $\gamma, m$ as in the exponential case for the quintessence and phantom fields, respectively.

\begin{figure*}[h!]
     \centering
     \begin{subfigure}[b]{0.49\textwidth}
         \centering
         \includegraphics[width=\textwidth]{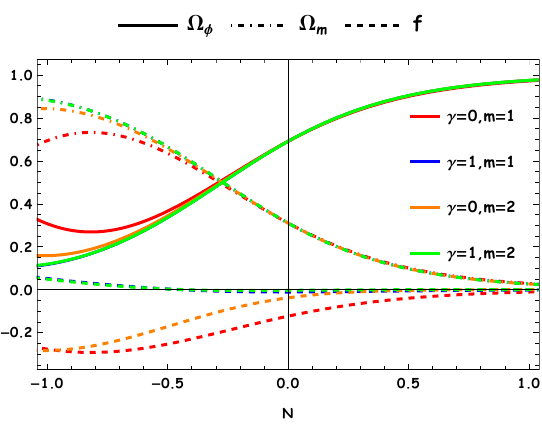}
         \caption{The plot of the evolution of the density parameter $\Omega_\phi$ (in solid) and $\Omega_m$ (in dot-dashed) and the interaction $f(x,y)$ (in dashed) of the quintessence field for $V(\phi) = (A + B \phi)^2$ .}
         \label{fig:nonexp_om_q}
     \end{subfigure}
     \hfill
     \begin{subfigure}[b]{0.49\textwidth}
         \centering
         \includegraphics[width=\textwidth]{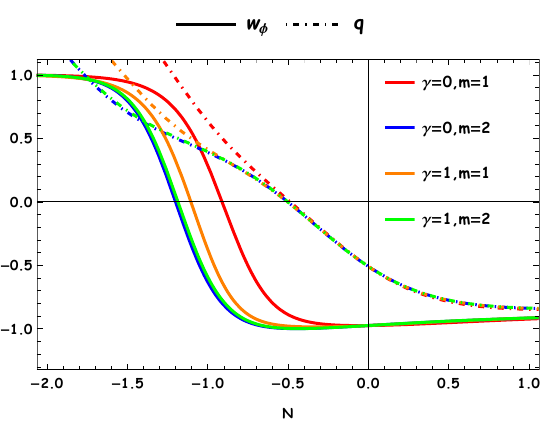}
         \caption{The plot of the evolution of the equation of state of the scalar field $w_\phi$ (in solid) and the deceleration parameter $q$ (in dashed) of the quintessence field for $V(\phi) = (A + B \phi)^2$.}
         \label{fig:nonexp_w_q}
     \end{subfigure}
     \hfill
      \begin{subfigure}[b]{0.49\textwidth}
         \centering
         \includegraphics[height=0.8\textwidth,width=1.\textwidth]{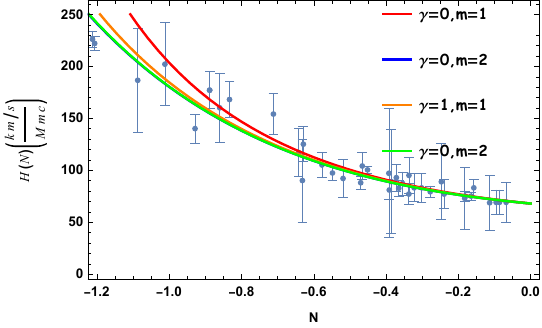}
         \caption{Plot of the $H(N)$ vs $N$ together with the observational data for the comparison of the quintessence field for $V(\phi) = (A + B \phi)^2$.}
         \label{fig:nonexp_H_q}
     \end{subfigure}
     \hfill
     \begin{subfigure}[b]{0.49\textwidth}
         \centering
         \includegraphics[width=\textwidth]{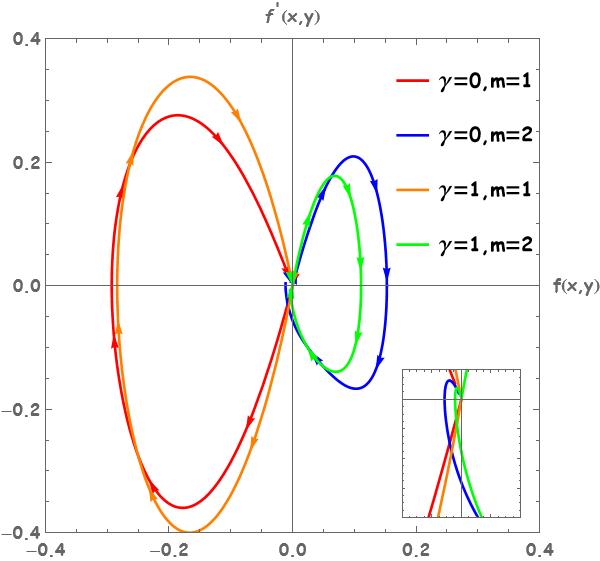}
         \caption{Phase plot displaying $f(x,y)$ against its derivative $f'(x,y)$ of the quintessence field for $V(\phi) = (A + B \phi)^2$. The inset showcases a zoomed-in view of the plot's central region for clarity.}
         \label{fig:nonexp_f_q}
     \end{subfigure}
        \caption{Plot of the different cosmological parameters and the phase space of the interaction $f(x,y)$ for the quintessence field for the $\Gamma=1/2$ corresponding to of the quintessence field for $V(\phi) = (A + B \phi)^2$ with $\alpha = -0.4$.}
        \label{fig:nonexp_quint}
\end{figure*}

In Fig.~\ref{fig:phase-q1}, we present a 3D phase plot of the quintessence model with $\Gamma = \frac{1}{2}, m=1$, $\gamma = [0,1]$ and $\alpha=-0.4$. The region that is permitted by the Friedmann constraint forms a cylindrical shape in the plot.
The cylinder's axis is aligned with the $\lambda$ axis. This is depicted as a shaded cylindrical region.
The surface of this cylinder indicates regions dominated entirely by dark energy. Conversely, the axis of the cylinder represents regions dominated solely by matter. It can be noticed for all the plots we have considered the solutions originated 
from the points $x=\pm1, y=0$ \footnote{Though the solutions originated 
from the dark energy \textbf{domination are not} expected from the physical point of view it is due to not considering the contribution of radiation in our analysis since we are interested in late-time dynamics. As in the early times, the contribution 
from the matter is negligible; the scalar field works as a proxy to the radiation to fulfill the Friedmann constraint equation $\Omega_m + \Omega_\phi =1$. This argument can be validated from the evolution of the EoS of the scalar field in 
Fig:\ref{fig:nonexp_w_q} where the EoS of the quintessence field remains positive during the early dark energy domination, hence unable to drive the acceleration of the universe. This has been reported before in \cite{Roy:2014hsa} .}  are 
attracted towards complete dark energy domination, where there could be a contribution from both the potential and kinetic parts of the field. The whole $\lambda$ axis represents the matter domination, and as expected from our analytical finding, it remains saddle for all cases.

In Fig.~ \ref{fig:phase-ph1}, we present the 3D phase plot for the phantom scenario with parameters $\Gamma = \frac{1}{2}$, $m=1$, and $\gamma = [0,1]$ for $\alpha = -0.4$. The blue-shaded hyperboloid regions represent the Friedmann constraint area for the phantom field. Compared to the quintessence case, the phase space for the phantom is more complex. Solutions can originate in the region where $\Omega_\phi < 0$, but eventually enter the Friedmann constraint region, being drawn towards late-time attractors dominated by the phantom field in the $\lambda = 0$ plane.

\begin{figure*}[h!]
     \centering
     \begin{subfigure}[b]{0.49\textwidth}
         \centering
         \includegraphics[width=\textwidth]{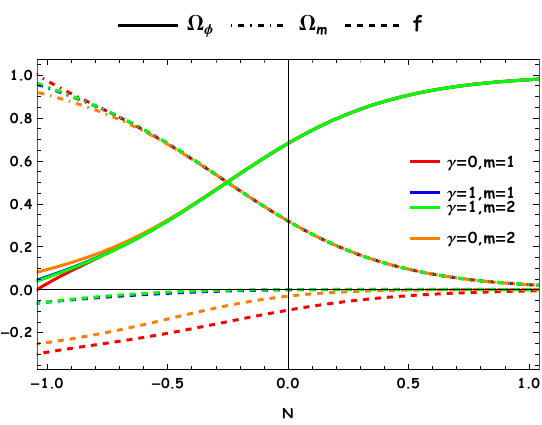}
         \caption{The plot of the evolution of the density parameter $\Omega_\phi$ (in solid) and $\Omega_m$ (in dot-dashed) and the interaction $f(x,y)$ (in dashed) of the phantom field for $V(\phi) = (A + B \phi)^2$.}
         \label{fig:nonexp_om_ph}
     \end{subfigure}
     \hfill
     \begin{subfigure}[b]{0.49\textwidth}
         \centering
         \includegraphics[width=\textwidth]{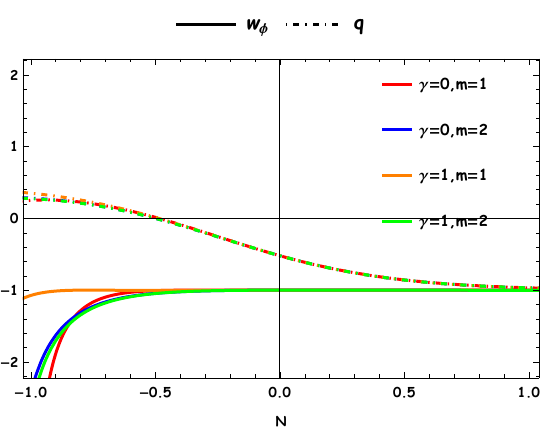}
         \caption{The plot of the evolution of the equation of state of the scalar field $w_\phi$ (in solid) and the deceleration parameter $q$ (in dashed) of the phantom field for $V(\phi) = (A + B \phi)^2$.}
         \label{fig:nonexp_w_ph}
     \end{subfigure}
     \hfill
      \begin{subfigure}[b]{0.49\textwidth}
         \centering
         \includegraphics[height=0.8\textwidth,width=1.\textwidth]{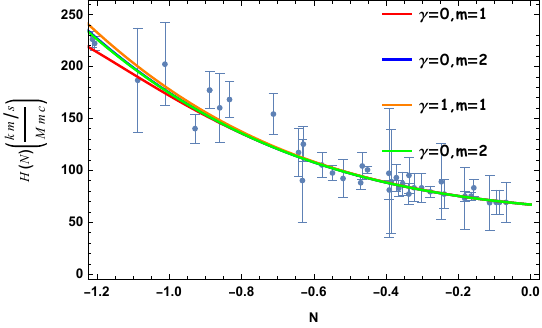}
         \caption{Plot of the $H(N)$ vs $N$ together with the observational data for the comparison of the phantom field for $V(\phi) = (A + B \phi)^2$.}
         \label{fig:nonexp_H_ph}
     \end{subfigure}
     \hfill
     \begin{subfigure}[b]{0.49\textwidth}
         \centering
         \includegraphics[width=\textwidth]{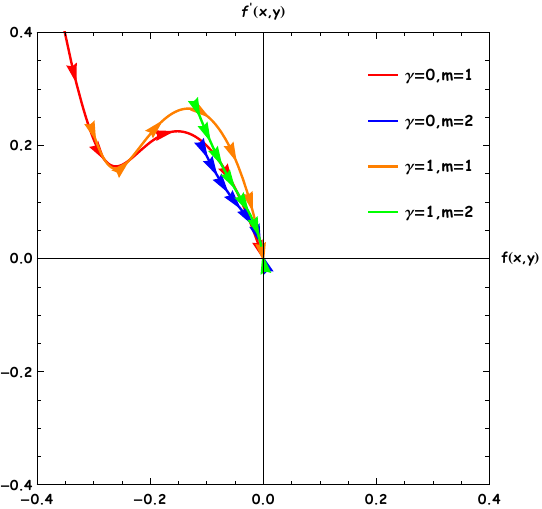}
         \caption{Phase plot displaying $f(x,y)$ against its derivative $f'(x,y)$ of the phantom field for $V(\phi) = (A + B \phi)^2$.}
         \label{fig:nonexp_f_ph}
     \end{subfigure}
        \caption{Plot of the different cosmological parameters and the phase space of the interaction $f(x,y)$ for the phantom field for the $\Gamma=1/2$ corresponding to of the quintessence field for $V(\phi) = (A + B \phi)^2$ with $\alpha=-0.4$.}
        \label{fig:nonexp_ph}
\end{figure*}

For further investigation of the system's evolution, we have considered the same strategy as the exponential case. The numerical simulation of the system has been considered with the estimated current values of $x_0, y_0$ as in the exponential case.  In Fig.~\ref{fig:nonexp_quint} and Fig.~\ref{fig:nonexp_ph}, we have shown the evolution of the cosmological parameters for the quintessence and phantom fields for the choice of $\Gamma = 1/2$ and $\alpha=-0.4$. For both the quintessence and the phantom case, the evolution of the energy densities $\Omega_\phi$ and $\Omega_m$ shows the expected behavior, and the interaction starts from a non-zero value and vanishes with time as the late time attractor is always dark energy dominated. The phase plot behavior of the interaction for the quintessence and the phantom field in Fig.~\ref{fig:nonexp_f_q} and Fig.~\ref{fig:nonexp_f_ph} show a behavior similar to the exponential potential counterpart. For the quintessence 
case, there can be a flip in the sign of the $f(x,y)$, indicating a change in the direction of the flow of energy from dark matter to dark energy. However, for the phantom case, it is unidirectional from dark matter to dark energy only.
For the quintessence field, $w_\phi >-1$ at an early time and approaches to  $w_\phi \simeq -1$ at late times and for the phantom field $w_\phi <-1$ at an early time and approaches to  $w_\phi \simeq -1$. It can also be also seen 
from the plots Fig.~\ref{fig:nonexp_H_q} and Fig.~\ref{fig:nonexp_H_ph} both that the quintessence and the phantom field with the nonexponential potentials can fit the  $H(z)$ vs $z$ data, particularly at late 
time. 

\section{Comparison with the existing results}

Some popular forms of the interaction term are given in Table \ref{tab:interactions}, which are all special cases of the present investigation.

A similar approach to ours has been taken up in \cite{Gumjudpai:2005ry} by considering both the quintessence and phantom fields in a single setup with an interaction of the type $Q = \beta \rho_m \dot{\phi}$ which is a subcase of our general interaction form (please see Table \ref{tab:interactions}, number 1). Our findings using the general setup are very similar to those in \cite{Gumjudpai:2005ry}. They indicate that the late-time attractors are fully dark energy dominated for both the quintessence and phantom fields, regardless of the form of the interaction.

In \cite{Chen:2008pz}, interactions of the forms $Q=\beta H \rho_m$ and $Q=\frac{\beta \rho_m \phi^2}{H}$ were examined for the quintessence field with exponential potentials. These correspond to subcases 2 and 4 in Table \ref{tab:interactions}, respectively. This study highlights the existence of a scaling attractor solution, which aligns with our analysis regarding the fixed point Q3 for the exponential potential case.

The interaction form $Q=\beta H {\dot{\phi}}^2$, as studied in \cite{Mimoso:2005bv} (subcase no 3), was analyzed for warm inflation with a generic potential. Their study found that the existence of a late-time scaling solution depends on both the asymptotic behaviour of the scalar field potential and the interaction. Similarly, in our analysis, extending this interaction to the late-time accelerated expansion of the universe reveals comparable behavior to the inflationary scenario. The stability of fixed points associated with the scaling regime explicitly relies on the asymptotic behavior of both the potentials and the interaction.

Subcase no. 5 from the Table \ref{tab:interactions} has been studied in \cite{Bernardi:2016xmb}, and it has been mentioned that to get a complete matter-dominated fixed point, the interaction must vanish. In our analysis, this is a general observation. For any arbitrary form of interaction term, that has to vanish for a complete matter-dominated fixed point, which is also justified from the physical point of view.

The novelty of our approach lies in investigating the dynamics of the IDE models from a broad perspective for both the quintessence and phantom fields. Not only the nature of the interaction but also the kind of potentials are chosen to be very general to start with. This generalized method can be particularly useful in comparing the IDE model against observational data, which will be our future endeavor. This approach may be applied to studying other dark energy models as well.

\section{Conclusions}

This study uses dynamical system analysis to investigate interacting dark energy models, including both quintessence and phantom scalar fields in a single setup through a switch parameter $\epsilon$. At first, the equations of motion of the scalar fields are recast to a set of autonomous systems by considering suitable variable transformations. Also, a general approach has been adopted for the choice of the potential. The choice of potential is classified into two general classes: exponential potential and non-exponential potential. Stability analysis has been performed without considering any particular form of the interaction, both for the exponential and non-exponential classes of the potentials.

A general form of interaction has been proposed as an example that can incorporate a wide class of popular forms of interaction. The numerical evolution of the system considering this form of interaction has been studied for both the exponential and non-exponential potentials. We have compared the evolution of the Hubble parameter in these models against the observed data, and it can be seen that these models can fit the data, very well if the parameters are chosen properly.

From our analysis, it is evident that the late-time attractor is a completely dark energy-dominated state of the universe. The numerical evolution of the universe suggests that there can be some interaction between the dark sectors, although this interaction becomes weaker with time and becomes negligible as the universe becomes more and more dark energy dominated. Our finding shows that for the quintessence field, the interaction can be from dark matter to dark energy and vice versa. During the evolution of the universe, the interaction might have started from dark matter to dark energy and it reversed its direction since there is a flip in the sign of the interaction term $f(x,y)$ and vanishes in the future. For the phantom field, the interaction is from dark matter to dark energy and vanishes over time. 

In conclusion, from our general qualitative analysis using a dynamical system approach, although the possibility of interaction between the dark sectors cannot be ruled out, there is no significantly compelling requirement of an interaction in the dark sectors either. Our future endeavour will be regarding the statistical test or preference of the models against recent cosmological data and also to check if the interactions can alleviate the current tensions in cosmology.

\begin{acknowledgements}
This work was supported by the Office of the Permanent Secretary, Ministry of Higher Education, Science, Research and Innovation (OPS MHESI), Thailand Science Research and Innovation (TSRI), and the University of Phayao (Grant No. RGNS 65-132). CK would like to acknowledge that this work was partially supported by the University of Phayao's International Sabbatical Leave Fund 2022. In addition, the authors would like to thank Burin Gumjudpai and Manabendra Sharma for the useful discussion on the work.
\end{acknowledgements}

\appendix
    
\section{Eigenvalues of the $Q_3$ fixed point.}\label{appen:eigenQ3}
The  trace and determinant of this fixed point are;


\begin{equation}
    \begin{split}
    T_{Q3}  &= \epsilon \partial_x f  +3 \sqrt{\frac{3}{2}} \lambda  x-6,\\
       D_{Q3}&= -3 x^2 \left(-5 \lambda ^2+M-3 \epsilon \right)\\
       &+\sqrt{6} \lambda  x \left(M \epsilon +\lambda ^2 (-\epsilon )-9\right) \\ 
       & -3 M \epsilon -N x \sqrt{9 x^2 \epsilon -3 \sqrt{6} \lambda  x+9}\\
       & +\frac{\lambda  N \epsilon  \sqrt{3 x^2 \epsilon -\sqrt{6} \lambda  x+3}}{\sqrt{2}} \\
       & -6 \sqrt{6} \lambda  x^3 \epsilon +3 \lambda ^2 \epsilon +9
    \end{split}
\end{equation}
The condition to have a stable region $T_{Q3}<0$ and $D_{Q3}>0$.

\section{Form of interactions}\label{appen:interactions}
In the following, we give a list of popular forms of interactions already studied in the literature that can be incorporated into our general parametrization of the interaction given in Eq.\ref{eq:int_example},

\begin{table}[h]
    \centering
    \begin{tabular}{|c|c|c|c|}
    \hline
    No & $Q$    & $q$ & References\\
     \hline
    1 &  $\beta \rho_m \dot{\phi}$   & $\frac{\sqrt{3}}{\sqrt{2}} \beta\left(1-x^2-y^2\right)$ & \cite{Amendola:1999qq,Gonzalez:2006cj,Boehmer:2008av,Gumjudpai:2005ry}\\
      \hline
       2 &  $
\beta H \rho_m $ & $ \frac{\beta}{2}\left(1-x^2-y^2\right) / x$  & \cite{Billyard:2000bh,Chen:2008pz}\\
         \hline
3 & $\beta H \dot{\phi}^2$  & $\beta x$ & \cite{Mimoso:2005bv,Roy:2018eug}\\
         \hline
    4 & $\beta \rho_m \dot{\phi}^2 / H$    & $3 \beta x\left(1-x^2-y^2\right)$ & \cite{Chen:2008pz}\\
         \hline   5 &   $\alpha\left(\rho_m +\rho_\phi\right) \dot{\phi}$   & $\alpha / \sqrt{6}$ & \cite{Bernardi:2016xmb}\\
         \hline
    \end{tabular}
    \caption{A list of example interactions that can be incorporated in the general form of the interaction considered in this work.}
    \label{tab:interactions}
\end{table}

\clearpage
\twocolumngrid
\bibliographystyle{unsrt}
\bibliography{qeos}

\end{document}